\documentclass[epj,twocolumn]{webofc}
\usepackage[varg]{txfonts}   
\woctitle{TRANSVERSITY 2014}
\begin{document}
\title{TMDs: Evolution, modeling, precision}
%
%

\newcommand{\al}{\alpha}
\newcommand{\eps}{\varepsilon}
\newcommand{\ga}{\gamma}
\newcommand{\Ga}{\Gamma}
\newcommand{\la}{\lambda}
\newcommand{\La}{\Lambda}
\newcommand{\nn}{\nonumber}
\newcommand{\ver}{\varepsilon_{\rm {IR}}}
\newcommand{\veu}{\varepsilon_{\rm {UV}}}
\newcommand{\bn}{{\bar n}}
\newcommand{\bv}{{\bar v}}
\newcommand{\pslash}{{\not \!p}}
\newcommand{\kslash}{{\not \!k}}
\newcommand{\nslash}{{\not \!n}}
\newcommand{\nbslash}{{\not \!\bn}}
\newcommand{\bnslash}{{\not \!\bn}}
\newcommand{\pin}{p_{in}}
\newcommand{\po}{p_{out}}
\newcommand{\nb}{\bar n}
\newcommand{\veir}{\varepsilon_{\rm {IR}}}
\newcommand{\veuv}{\varepsilon_{\rm {UV}}}
\newcommand{\GeV}{\rm GeV}

\newcommand{\lc}{\lowercase}
\newcommand{\be}{\begin{equation}}
\newcommand{\ee}{\end{equation}}
\newcommand{\bea}{\begin{eqnarray}}
\newcommand{\eea}{\end{eqnarray}}
\newcommand{\balign}{\begin{align}}
\newcommand{\ealign}{\end{align}}
\newcommand{\as}{\alpha_s}

\newcommand{\bnp}{\bar n \!\cdot\! p}
\newcommand{\bnP}{\bar {\cal P}}
\newcommand{\ppP}{{\cal P}_\perp}
\newcommand{\bnPd}{\bar {\cal P}^{\raisebox{0.8mm}{\scriptsize$\dagger$}} }
\newcommand{\cPslash}{ {\cal P}\!\!\!\!\slash}
\newcommand{\bs}{\!\hspace{0.05cm}}
\newcommand{\Ub}{{\cal U}}
\newcommand{\cD}{{\cal D}}
\newcommand{\AL}{A_X}
\newcommand{\DL}{D_X}
\newcommand{\WL}{W_X}
\newcommand{\SL}{S_X}
\newcommand{\WLd}{W_X^\dagger}
\newcommand{\SLd}{S_X^\dagger}

\newcommand{\pathexp}[1]{\exp \l( \frac{i}{\hbar} #1 \r) }
\newcommand{\bra}[1]{\left< #1 \right |}
\newcommand{\ket}[1]{\left | #1 \right >}
\newcommand{\braket}[2]{\left< #1 | #2 \right>}
\newcommand{\sandwich}[3]{\left< #1 \right | #2 \left | #3 \right >}

\newcommand{\cc}{\cite}
\newcommand{\bg}{\begin{gather}}
\newcommand{\foma}{\end{gather}}

\newcommand{\noopsort}[1]{} \newcommand{\printfirst}[2]{#1}
\newcommand{\singleletter}[1]{#1} \newcommand{\switchargs}[2]{#2#1}

\newcommand{\Ree}{\mbox{Re}}
\newcommand{\Imm}{\mbox{Im}}
\newcommand{\vecb}[1]{\mbox{\boldmath $#1$}}
\newcommand{\vecbe}[1]{\mbox{\boldmath ${\scriptstyle #1}$}}
\newcommand{\vecbp}[1]{\mbox{\boldmath $#1_\perp$}}\newcommand{\matrm}[1]{\mbox{\scriptsize #1}}
\newcommand{\matr}[1]{\mbox{#1}}

\def\halb{\frac{1}{2}}
\def\e{\epsilon}
\def\E{\hbox{$\cal E $}}
\def\ort{\hbox{$\cal T$}}
\def\lagr{\hbox{$\cal L$}}
\def\ve{\varepsilon}
\def\w{\omega}
\def\hb{\hbar}
\def\pd{\partial}
\def\f{\phi}
\def\F{\Phi}
\def\L{\Lambda}
\def\pb{\bar \psi}
\def\W{\Omega}
\def\z{\zeta}
\def\M{\bar\m}
\def\ex{\hbox{e}}
\def\S{\Sigma}
\def\F{\Phi}
\def\<{\langle}
\def\>{\rangle}
\def\th{\tanh}
\def\ch{\cosh}
\def\sh{\sinh}
\def\sign{\hbox{sign}}
\def\arcth{\hbox{arcth}}
\def\h{\hbar}
\def\a{\alpha}
\def\b{\beta}
\def\g{\gamma}  \def\G{\Gamma}
\def\d{\delta}  \def\D{\Delta}
\def\l{\lambda}   \def\L{\Lambda}
\def\s{\sigma}
\def\r{\rho}  \def\vr{\varrho}
\def\x{\xi}
\def\c{\chi}
\def\m{\mu}
\def\n{\nu}
\def\t{\tau}
\def\k{\kappa}
\def\z{\zeta}
\def\w{\omega}
\def\vf{\varphi}
\def\({\left(}
\def\[{\left[}
\def\){\right)}
\def\]{\right]}
\def\coth{\hbox{coth}}
\def\cot{\hbox{cot}}
\def\cos{\hbox{cos}}
\def\sin{\hbox{sin}}
\def\ln{\hbox{ln}}
\def\log{\hbox{log}}
\def\inf{\infty}
\def\dk{{d^n k \over (2\pi)^n}}
\def\Tr{\hbox{Tr}}
\def\pa{{\cal P}}
\def\dprop{D_{\m\n}}
\def\sivf{f_{1T}^{\a \perp}(x, {k}_{\perp}^2)}
\def\ndown{\underline{\vecc n}}
\def\nup{\bar{\vecc n}}
\def\nbar{\bar{n}}

\def\Slash#1{{#1\!\!\!\slash}}

\def\Qslash{Q\!\!\!\!\slash}
\def\Dslash{D\!\!\!\!\slash}
\def\ppslash{p^{\,\prime}\!\!\!\!\!\slash}
\def\nslash{n\!\!\!\slash}
\def\bnslash{\bar n\!\!\!\slash}
\def\pslash{p\!\!\!\slash}
\def\bpslash{\bar p\!\!\!\slash}
\def\qslash{q\!\!\!\slash}
\def\kslash{k\!\!\!\slash}
\def\lslash{l\!\!\!\slash}
\def\vslash{v\!\!\!\slash}
\def\pdslash{\partial\!\!\!\slash}
\def\Aslash{A\!\!\!\slash}
\def\SppP{{\cal {P\!\!\!\!\hspace{0.04cm}\slash}}_\perp}

\def \le { \left    }
\def \ri { \right }

\def\bp{\bar p}
\def\bP{\bar P}

\def\lqcd{\L_{\rm QCD}}

\newcommand{\blue}[1]{{\color{blue} {\bf #1}}}
\newcommand{\red}[1]{{\color{red} {\bf #1}}}
\newcommand{\magenta}[1]{{\color{magenta} {\bf #1}}}
\newcommand{\orange}[1]{{\color{YellowOrange} {\bf #1}}}


\author{Umberto D'Alesio,  \inst{1}\fnsep\thanks{\email{umberto.dalesio@ca.infn.it}} \and
        Miguel G. Echevarr\'ia,
\inst{2}\fnsep\thanks{\email{m.g.echevarria@nikhef.nl}} \and
        Stefano Melis\inst{3}\fnsep\thanks{\email{stefano.melis@to.infn.it}}
\and
        Ignazio Scimemi\inst{3}\fnsep\thanks{Speaker, \email{ignazios@fis.ucm.es}}
}

\institute{Dipartimento di Fisica, Universit\`a di Cagliari, and INFN, Sezione di Cagliari, Cittadella Universitaria di Monserrato, I-09042 Monserrato (CA), Italy
\and
NIKHEF and Department of Physics and Astronomy,
VU University Amsterdam, De Boelelaan 1081, NL-1081 HV Amsterdam, the Netherlands
\and
Dipartimento di Fisica, Universit\`a di Torino, Via P. Giuria 1, I-10125 Torino, Italy
\and
Departamento de F\'isica Te\'orica II, Universidad  Complutense de Madrid, 28040 Madrid, Spain
          }

\abstract{%
The factorization theorem for $q_T$ spectra in Drell-Yan processes, boson production and semi-inclusive deep inelastic scattering allows for the determination of the non-perturbative parts of transverse momentum dependent parton distribution functions. Here we discuss the fit of Drell-Yan and $Z$-production data using the transverse momentum dependent formalism and the resummation of the  evolution kernel.
We find a good theoretical stability of the results and a final $\chi^2/{\rm points}\lesssim 1$. We show how the fixing of the non-perturbative pieces of the evolution can be used to make predictions at present and future colliders.
}
\maketitle
\section{Introduction}
\label{intro}
The study of differential cross sections is notoriously a great source of information on the nature of fundamental interactions. Recently the factorization theorem for transverse momentum dependent cross sections formulated by two groups~\cite{GarciaEchevarria:2011rb,Echevarria:2012js,Echevarria:2014rua,Collins:2011zzd} has pointed out that in Drell-Yan (DY), semi-inclusive deep inelastic scattering (SIDIS) and $e^+e^-\rightarrow 2\; {\rm hadrons/jets}$ at high boson invariant mass, all non-perturbative QCD effects can be encoded in the so called Transverse Momentum Distributions (TMDs) and can be included in experiments run at different energies solving appropriate evolution equations. The evolution factors so derived are fixed by perturbative QCD only up to a certain level of accuracy, depending, among the others, on the initial and final center of mass energy scales. The fundamental issue behind the evolution between two scales of the TMDs is that the factorization theorem is valid in both energy regimes.

The object of this talk concerns the study  of TMD for initial states, the so called transverse momentum dependent parton distribution functions (TMDPDFs). As a first we want to study up to which level the evolution of TMDPDF can be fixed just using resummations of the perturbative series using the data of Drell-Yan and vector-boson production at hadron colliders currently available. This analysis illustrates some important points when comparing to other attempts to include non-perturbative QCD effects in differential cross sections, like in Ref.~\cite{Landry:2002ix,Konychev:2005iy,Echevarria:2014xaa,Aidala:2014hva}, and the use of non-perturbative models.
Finally we show the precision that can be achieved making predictions for some observables at the Large Hadron Collider (LHC). In particular we study the differential cross section for $Z$-boson production at the peak of the distribution.

The cross sections that we consider in this work can be formulated schematically according to the factorization formula~\cite{GarciaEchevarria:2011rb,Echevarria:2012js,Collins:2011zzd}
\begin{align}
\frac{d\sigma}{dq_T} & \sim
H(Q^2,\m^2)
\int d^2\vecb k_{AT}\, d^2\vecb k_{BT}  \d^{(2)}(\vecb k_{AT}+\vecb k_{BT}-\vecb q_{T})\, \nn \\ &
\times F_A(x_A,\vecb k_{AT};\z_A,\m)\,
F_B(x_B,\vecb k_{BT};\z_B,\m)\,
\,,
\label{eq:cs}
\end{align}
where $F_{A,B}$ are the TMDPDFs.
They depend on the dilepton invariant mass trough the scales $\z_A$ and $\z_B$\footnote{In Ref.~\cite{Echevarria:2012js,Echevarria:2014rua} the authors used the equivalent notation $\z_A=Q^2/\alpha$ and $\z_B=Q^2\alpha$, where $\alpha$ is the soft function splitting parameter.}, being $\z_A\z_B=Q^4$, the intrinsic parton transverse momenta, the factorization scale $\m$ and the lightcone momentum fractions.
Finally, $H$ is the hard factor, which is spin independent and can be calculated adopting the standard perturbation theory.

\section{Construction of TMDPDF}
 The construction of the TMDPDF which  are part of the cross section follows several steps, which can be found in  Ref.~\cite{D'Alesio:2014vja}  and we partially report here.

Parametrizing the non-perturbative large-$b_T$ region of the quark TMDPDF (similar expressions hold for the gluon TMDPDF), we write it at some initial scale $Q_i$ as
\begin{align}
\label{eq:FqN1}
\tilde F_{q/N}(x,b_T;Q_i^2,\m_i) =
{\tilde F}^{\rm pert}_{q/N}(x,b_T;Q_i^2,\m_i)\,
{\tilde F}^{\rm NP}_{q/N}(x,b_T;Q_i)
\,,
\end{align}
where ${\tilde F}^{\rm NP}_{q/N}(x,b_T;Q_i)$ is the  non-perturbative part of the TMDPDF with
\begin{align}
\label{eq:fgfg}
{\tilde F}^{\rm NP}_{q/N}(x,b_T;Q_i)\equiv
{\tilde F}^{\rm NP}_{q/N}(x,b_T)\left(\frac{Q_i^2}{Q_0^2}\right)^{-D^{\rm NP}(b_T)}
\,.
\end{align}
Notice that in the equation above we have parametrized the non-perturbative contribution in the same way as we do for the evolution kernel  that we describe  in Ref.~\cite{Echevarria:2012pw,D'Alesio:2014vja}.

In the cross section, Eq.~(\ref{eq:cs}), we fix the factorization scale $\m=Q$, so that we can write the resummed TMDPDF that enters into the factorization theorem as
\begin{align}
\label{eq:FRF2}&
\tilde F_{q/N}(x,b_T;Q^2,Q)=
\tilde R^{\rm pert}(b_T;(Q_0+q_T)^2,Q_0+q_T,Q^2,Q)\, \nn\\ &\times
{\tilde F}^{\rm pert}_{q/N}(x,b_T;(Q_0+q_T)^2,Q_0+q_T)\,
{\tilde F}^{\rm NP}_{q/N}(x,b_T;Q)
\,.
\end{align}
 The evolution   kernel $\tilde R(b_T;(Q_0+q_T)^2,Q_0+q_T,Q^2,Q)$ is here split in a perturbative calculable part, $\tilde R^{\rm pert}(b_T;(Q_0+q_T)^2,Q_0+q_T,Q^2,Q)$ and a non-perturbative piece which is included in  $ {\tilde F}^{\rm NP}_{q/N}(x,b_T;Q)$,
\begin{align}
\label{eq:FNP4}
{\tilde F}^{\rm NP}_{q/N}(x,b_T;Q)=
{\tilde F}^{\rm NP}_{q/N}(x,b_T;Q_i) \left(\frac{Q^2}{Q_i^2}\right)^{-D^{\rm NP}(b_T)}\ .
\end{align}
More explicitly, the TMDPDF is implemented as
\begin{align}
\label{eq:Ffull}
&\tilde F_{q/N}(x,b_T;Q^2,Q)
=
\exp\le\{
\int_{Q_i}^{Q} \frac{d\bar\m}{\bar\m}
\g_F\le(\as(\bar\m),\ln\frac{Q^2}{\bar\m^2} \ri)\ri\}\nn\\ &\times
\left(\frac{Q^2 b_T^2}{4 e^{-2\g_E} }\right)^{-D^R(b_T;Q_i)}
e^{h_\G^R(b_T;Q_i)-h_\g^R(b_T;Q_i)}\nn\\ &\times
\sum_j
\int_{x}^{1}\frac{dz}{z}
\hat C_{q\leftarrow j}(x/z,b_T;Q_i)\, f_{j/N}(z;Q_i)\,
{\tilde F}^{\rm NP}_{q/N}(x,b_T;Q)
\,,
\end{align}
where, as we already mentioned, $Q_i=Q_0+q_T$.
The are several points to be emphasized in this formula.
In order to minimize the value of the logarithms we choose $\m_i=Q_i$.
Next we notice that the splitting into a coefficient and a collinear parton distribution function (PDF) is valid only at high transverse momentum, so that we expect that the choice $Q_i= Q_0+q_T$ (where $Q_0$ is a fixed low scale) minimizes the logarithms generated by this splitting.
The scale $Q_0$ works as a minimum matching scale between the TMDPDF and the PDF, such that it sits at the border between the perturbative and non-perturbative regimes; in particular we choose $Q_0\sim 2~\GeV$.
The exponentiated pieces in Eq.~(\ref{eq:Ffull}), namely the factor
$\left(\frac{Q^2 b_T^2}{4 e^{-2\g_E} }\right)^{-D^R(b_T;Q_i)}$ and $
e^{h_\G^R(b_T;Q_i)-h_\g^R(b_T;Q_i)}
$ have their origin respectively in the expression of the evolution kernel  ~\cite{GarciaEchevarria:2011rb,Echevarria:2012js,Echevarria:2014rua,Collins:2011zzd} and in
an exponentiable part of the matching coefficient  between TMDPDF and
 PDF~\cite{Kodaira:1981nh,Becher:2011xn,Ceccopieri:2014qha}. These pieces can be further resummed  using the counting $\alpha_s(q_T) \ln  (b^2 q_T^2)\sim 1$ which is the relevant one for the small-$q_T$
region~\cite{Echevarria:2012pw,D'Alesio:2014vja}.

In order to fix the arguments of the non-perturbative part ${\tilde F}^{\rm NP}$, we need to consider the following constraints:
\begin{itemize}
\item It must correct the behavior of ${\tilde F}^{\rm pert}_{q/N}$ at large values of $b_T$, where the perturbative expansion looses its convergence properties and the Landau pole singularity shows up, both in the evolution kernel and in the matching coefficient of the TMDPDF onto the PDF.
\item It has to be such that
\begin{align}
\label{eq:blimit}
\lim_{b_T\rightarrow 0}{\tilde F}^{\rm NP}_{q/N}=1\,,
\end{align}
in order to guarantee that the perturbative series is not altered where its convergence properties are sound.
\end{itemize}
We have not included a dependence on $x$, as data eventually do not need such correction and to keep the model simple enough.
In Eq.~(\ref{eq:blimit}) we are assuming that the values of $x$ are not extremely small (say $x>10^{-3}$), in which case the whole TMD formalism should be re-considered.

We have studied several parametrizations of the non-perturbative part (Gaussian, polynomial, etc.) and the final one which better provides a good fit of the data, with the minimum set of parameters,  $D^{\rm NP}=0$, is
\begin{align}
\label{eq:FqN3}
{\tilde F}^{\rm NP}_{q/N}(x,b_T;Q) &=
e^{-\l_1 b_T}\le(1+\l_2 b_T^2\ri)
\,.
\end{align}
The data for $Z$-boson production are basically sensitive just to the parameter $\l_1$, that is to the exponential factor and not to the second power-like term.
The global fit so performed allows to fix, to a certain precision, the value of this non-perturbative constant. In other words, this fit can be used to fix the amount of non-perturbative QCD corrections in the transverse momentum spectra.
The parameter $\l_2$ corrects the behavior of the TMDPDF at high values of $b_T$ and results necessary to describe the data at low dilepton invariant mass and low $q_T$. The results of the fit for this case are shown in Tab.~\ref{tab:tevlow_CTEQ10_q0pt_param} and discussed  later in the text.

Considering now a nonzero $D^{\rm NP}$, this results in a $Q$-dependent factor in the non-perturbative model (see the studies of Refs.~\cite{Korchemsky:1994is,Tafat:2001in} and more recently Refs.~\cite{Collins:2011zzd,Echevarria:2014rua}).
Thus, from Eqs.~(\ref{eq:FNP4}) and (\ref{eq:FqN3}), by setting $D^{\rm NP}=\l_3 b_T^2/2$, we have
\begin{align}
\label{eq:FqN3m}
{\tilde F}^{\rm NP}_{q/N}(x,b_T;Q)
&=
e^{-\l_1 b_T}\le(1+\l_2 b_T^2\ri)\le(\frac{Q^2}{Q_0^2}\ri)^{-\frac{\l_3}{2} b_T^2}\,.
\end{align}

The sensitivity of the data to this extra factor involving $\l_3$ is not very  strong, although we observe an improvement in the $\chi^2$.
This is a consequence of the fact that the fully resummed $D$ function is actually valid in a region of impact parameter space which is broad enough for the analysis of the sets of available data (notice that we have, in all cases, a dilepton invariant mass $Q>4$ GeV). It might be that at lower values of $Q$ such corrections could be more significant. On the other hand one expects that also the factorization theorem should be revised when the values of $Q$ become of the order of the hadronization scale. It is then possible that the non-perturbative corrections to the evolution kernel happen there where the basic hypothesis of the factorization theorem ($Q\gg q_T\sim\Lambda_{QCD}\sim {\cal O}(1\;{\rm GeV})$) become weaker and so are more difficult to extract. A more detailed study in this direction is beyond the scope of this paper. The results of the fit for the model in Eq.~(\ref{eq:FqN3m}) are shown in Tab.~\ref{tab:CTEQ}  and Fig.~\ref{fig:l3} and discussed  later in the text.

As a general remark one has to keep in mind that in practical calculations we have eventually to Fourier transform the product of two TMDPDFs. The integration in impact parameter space is done numerically over a suitable $b_T$ range. We have checked that the region outside the endpoints of this integration does not affect the final result. In fact, the points for very small $b_T$ are relevant only for extremely high transverse momenta, which is not the case in our study. At very high $b_T$ the TMDs are completely negligible.

To conclude this section we observe that while the parameter $\la_3$, being a correction to the $Q$-dependent piece of the TMD, is flavor independent, the other  parameters $\la_{1,2}$ can in principle be flavor dependent.
In the fit that we have performed we have not included this feature, namely for two reasons: $i)$ the DY data that we use depend just on one combination of $\l_{1,2}$ (remember that we consider neutral current mediated processes and only nucleons as initial states); $ii)$ the quality of the fit is so good that we would not be sensitive (statistically) to the flavor dependence of these parameters.
Nevertheless the inclusion of data from processes with different initial states and/or mediated by charged currents could definitely help in this respect.

\section{Data selection}
The factorization theorem for Drell-Yan and vector boson production has its own range of validity. The main condition  is that invariant mass, $Q$, is much bigger than the hadronization scale, $\sim 1$ GeV,
 and the transverse momenta involved.  This is actually the case for the data that we have considered
~\cite{Ito:1980ev,Moreno:1990sf,Antreasyan:1981uv,Affolder:1999jh,Abbott:1999yd,Abbott:1999wk,
Aaltonen:2012fi,Abazov:2007ac}. The  bin with the lowest  center of mass energy has $Q\sim 4$ GeV and $q_T\lesssim 1.4$ GeV.

Definitely we need data at moderate center of mass energies covering the small-$q_T$ region (up to 1-2 GeV) and intermediate dilepton invariant mass values (below 10 GeV). These come mainly from fixed-target experiments.  On the other hand to access larger $q_T$ values and even larger scales we have to include also high-energy collider experimental data, like those from Tevatron at the $Z$-boson peak. In both cases we will keep fulfilling the requirement $q_T\ll Q$, region of application of our approach.
Notice that to conform with the standard notation adopted in experimental analysis in the following we use $M=Q$ for the dilepton invariant mass.

These two classes of data are indeed complementary and essential to test the scale evolution of TMDs over a suitable range of scale values and to quantify the role of the non-perturbative part entering these distributions.

While for the low-energy data we consider the invariant differential cross section in the virtual boson momentum, for the high-energy data sets we use the ratio of the their $q_T$ dilepton distribution normalized to the experimental total cross section. In such a case, we compute this numerator following our approach and use the normalization factor as obtained with the DYNNLO code of Catani \emph{et al.}~\cite{Catani:2007vq,Catani:2009sm}. The use of this ratio avoids the problem of the discrepancy between D0 and CDF experimental results that could cause a source of systematics and/or tension between data sets.

We perform a fit both at next-to-leading-logarithm (NLL) accuracy as well as at next-to-NLL (NNLL). When adopting the NLL approximation we use the next-to-leading-order (NLO) collinear parton distributions, while at NNLL we use the next-to-NLO (NNLO) PDFs. In both cases we adopt $Q_i=Q_0+q_T$ . For the collinear PDFs we have tested both the MSTW08~\cite{Martin:2009iq} and the CTEQ10~\cite{Lai:2010vv} sets and we find a complete consistency among the results.

\begin{figure}[h]
 \begin{center}
\hspace{-2cm}
\includegraphics[width=0.5\textwidth, angle=0,natwidth=610,natheight=642]
{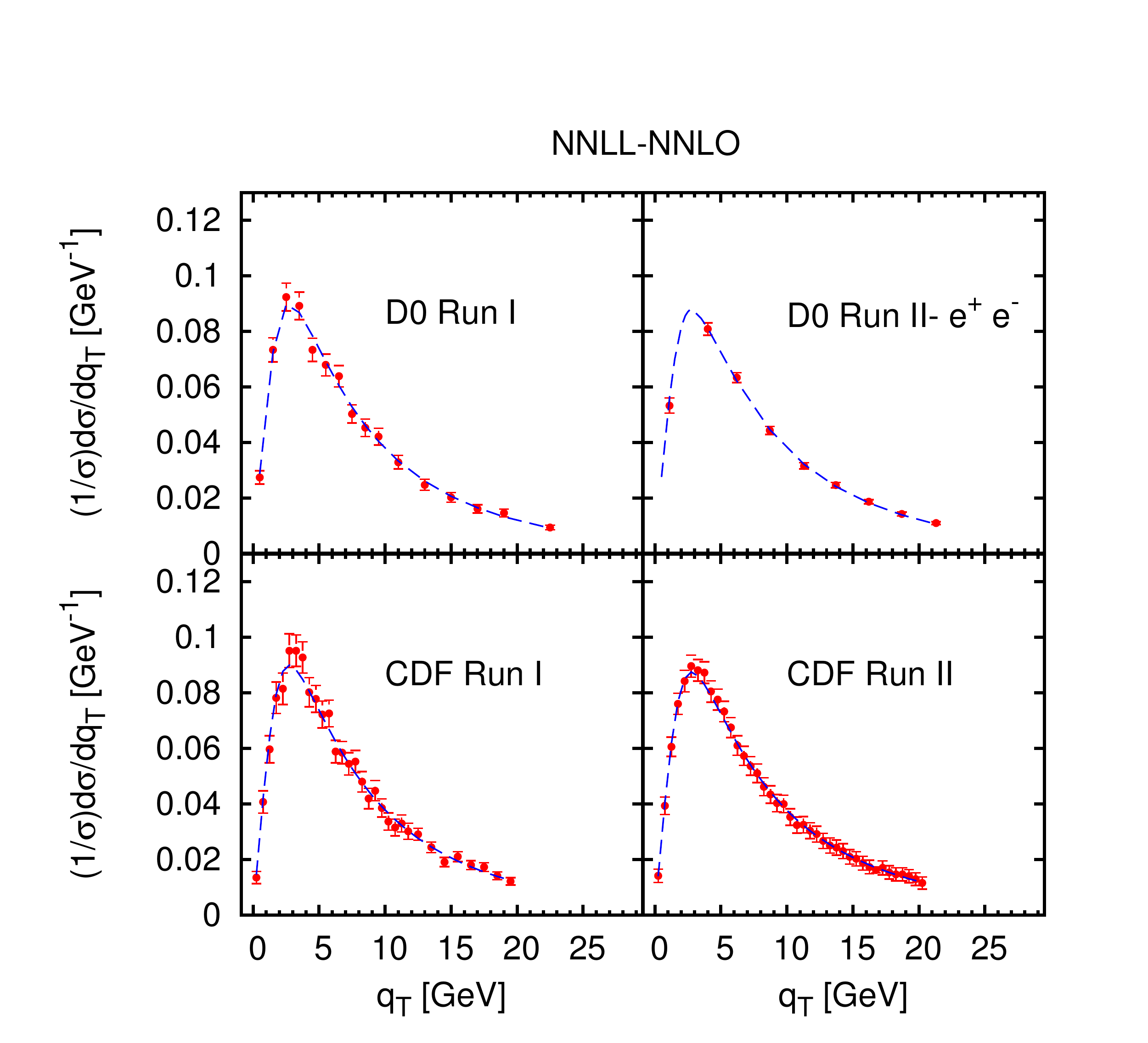} \vspace*{-.1cm}
\hspace{-1.5cm}
 \includegraphics[width=.55\textwidth, angle=0,natwidth=610,natheight=642]{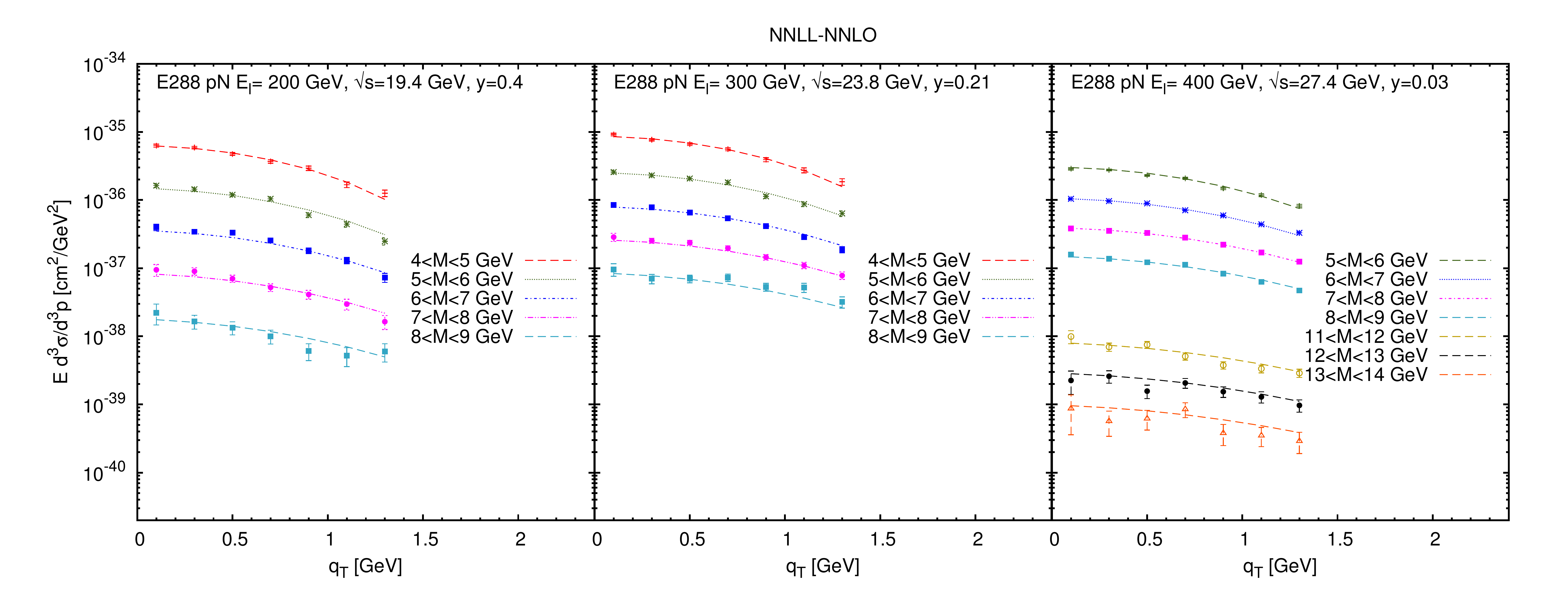}
 \caption{Best-fit curves for the analysis with $D^{\rm NP}\ne0$ (Eq.~(\ref{eq:FqN3m})), at NNLL accuracy
using the MSTW08~\cite{Martin:2009iq} set for the PDFs. Comparison with Tevatron data (upper-left panel), with R209 data (upper-right panel) and E288 data (lower panels).
\label{fig:l3}}
 \end{center}
 \end{figure}

One of the main goals of this work consists in the fits performed at NNLL accuracy with full resummation.
The NLL fits are mainly used as a check of convergence of the theory and other phenomenological aspects. We have tested both a $Q$-independent and $Q$-dependent parametrization of the non-perturbative inputs as given respectively in Eq.~(\ref{eq:FqN3}) and Eq.~(\ref{eq:FqN3m}).  The  two models give consistent results and the model of Eq.~(\ref{eq:FqN3m}) shows a slight better description for  Drell-Yan data. Some results are reported  in Tab.~\ref{tab:tevlow_CTEQ10_q0pt_param}-\ref{tab:CTEQ}, where we have used the CTEQ10~\cite{Lai:2010vv}  for the PDFs. In
Fig.~\ref{fig:l3} we show the results of the fit at NNLL using the MSTW08~\cite{Martin:2009iq} set for the PDFs. In these tables we report just the statistical error on the fitted parameters. The theoretical error, which includes the scale variation and  the error  due to the choice of PDF,  is of the same order as  the statistical error  for the NLL analysis and much smaller then the statistical error in the case of the  NNLL fit.
The theoretical errors
 (which include the error due to uncalculated perturbative terms still using the full resummation),
is estimated studying the dependence on the initial scale $Q_i=Q_0+q_T$ (where $Q_0=2$ GeV) in two ways: $i)$ we check the impact of a change in $Q_0$ allowing $m_{\rm charm}\sim  1.3 \;{\rm GeV}\leq Q_0\leq 2.7$ GeV, where the lowest value of $Q_0$ is about the charm threshold and the highest value is limited by the energy of the lowest energy bin of data; $ii)$ keeping  $Q_0=2$ GeV, we vary $Q_0+q_T/2\leq Q_i\leq {\rm min}\;(Q_0+2 q_T, Q)$. In the first case the fit is practically unaffected concerning the values of the parameters $\l_1,\l_2$.
For the second case the scale dependence instead has some impact on the these values. In particular  at NLL the theoretical error is of the same order of the statistical one and there is a clear reduction of the scale dependence at NNLL.
At this order the main uncertainty on the fitted parameters comes from the statistical error.
The statistical error is estimated requiring a 68\% confidence level, corresponding to a $\Delta \chi^2=4.72$ for four parameters.

\begin{table}[h]
\renewcommand{\tabcolsep}{0.4pc} 
\renewcommand{\arraystretch}{1.2} 
\begin{center}
\begin{tabular}{|l|l|}
 \hline
          NLL                       &              \\
  \hline
 223  points                                  &  $\lambda_1$ = $0.28\pm0.05_{stat}\textrm{ GeV}$
\\
&   $\lambda_2=0.14\pm0.04_{stat}\textrm{ GeV}^2$  \\
  $\chi^2$/dof = 1.79                                   &  $N_{\rm E288}=1.02\pm0.04_{stat}$
\\
&$N_{\rm R209}=1.4\pm0.2_{stat}$\\
\hline
\hline
          NNLL                       &             \\
  \hline
   223  points                                  &  $\lambda_1$ = $0.32\pm0.05_{stat}\textrm{ GeV}$
\\ &   $\lambda_2=0.12\pm0.03_{stat}\textrm{ GeV}^2$  \\
$\chi^2$/dof = 0.96                                     &  $N_{\rm E288}=0.99\pm0.05_{stat}$
\\
&$N_{\rm R209}=1.6\pm0.3_{stat}$\\
\hline
\end{tabular}
\end{center}
\caption{
Results of our global fit on low-energy~\cite{Ito:1980ev,Antreasyan:1981uv} and Tevatron data~\cite{Affolder:1999jh,Abbott:1999yd,Abbott:1999wk, Aaltonen:2012fi, Abazov:2007ac}, with $D^{\rm NP}=0$ (Eq.~(\ref{eq:FqN3})), \mbox{$Q_i=Q_0+q_T$}, at NNLL and NNL accuracies and with the collinear parton distributions from   CTEQ10~\cite{Lai:2010vv}
 at NNLO and NLO.
\label{tab:tevlow_CTEQ10_q0pt_param}}
\end{table}

\begin{table}[h]
\vskip 18pt
\renewcommand{\tabcolsep}{0.4pc} 
\renewcommand{\arraystretch}{1.2} 
\begin{tabular}{|c|c|c|c|}
 \hline
 &     NNLL               &           NLL           \\
 \hline
 \hline
   $\lambda_1$       & $0.29\pm0.04_{stat}\textrm{ GeV}$   & $0.27\pm0.06_{stat}\textrm{ GeV}$    \\
  \hline
   $\lambda_2$       & $0.170\pm0.003_{stat}\textrm{ GeV}^2$ & $0.19\pm0.06_{stat}\textrm{ GeV}^2$   \\
  \hline
    $\lambda_3$       & $0.030\pm0.01_{stat}\textrm{ GeV}^2$ & $0.02\pm0.01_{stat}\textrm{ GeV}^2$   \\
  \hline
  $N_{E288}$    &    $0.93\pm0.01_{stat}$              &  $0.98\pm0.06_{stat}$                              \\
 \hline
    $N_{R209}$  &       $1.5\pm0.1_{stat}$             &  $1.3\pm0.2_{stat}$                                \\
  \hline
   $\chi^2$              &       180.1               &    375.2                                         \\
  \hline
  \hline
  points                  &  $\chi^2/\textrm{points}$ &  $\chi^2/\textrm{points}$            \\
  \hline
     223                 &      0.81                 &     1.68                              \\
 \hline
  ~            &  $\chi^2/\textrm{dof}$ &  $\chi^2/\textrm{dof}$            \\
 \hline
                        &      0.83                 &     1.72                               \\
  \hline
  \hline
  E288 200              &       1.35                 &     2.28                            \\
  \hline
 E288 300            &       0.98                 &     1.22                          \\
 \hline
 E288 400            &       1.05                 &     2.33                           \\
 \hline
 \hline
 R209            &       0.27                 &     0.40                            \\
 \hline
 \hline
  CDF Run I             &       0.70                 &     1.50                              \\
 \hline
  D0 Run I           &       0.41                 &     1.77                                \\
 \hline
  CDF Run II            &       0.25                 &     0.76                            \\
 \hline
  D0 Run II           &       0.82                 &     3.2                          \\
 \hline
 \end{tabular}
 \caption{
Results of our global fit on low-energy~\cite{Ito:1980ev,Antreasyan:1981uv} and Tevatron data~\cite{Affolder:1999jh,Abbott:1999yd,Abbott:1999wk, Aaltonen:2012fi, Abazov:2007ac}, with $D^{\rm NP}\ne0$ (Eq.~(\ref{eq:FqN3m})), \mbox{$Q_i=Q_0+q_T$}, at NNLL and NNL accuracies and with the collinear parton distributions from CTEQ10~\cite{Lai:2010vv} at NNLO and NLO.
\label{tab:CTEQ}
 }
\end{table}

\begin{figure}
\centering
\hspace{-2cm}
\includegraphics[width=10cm]{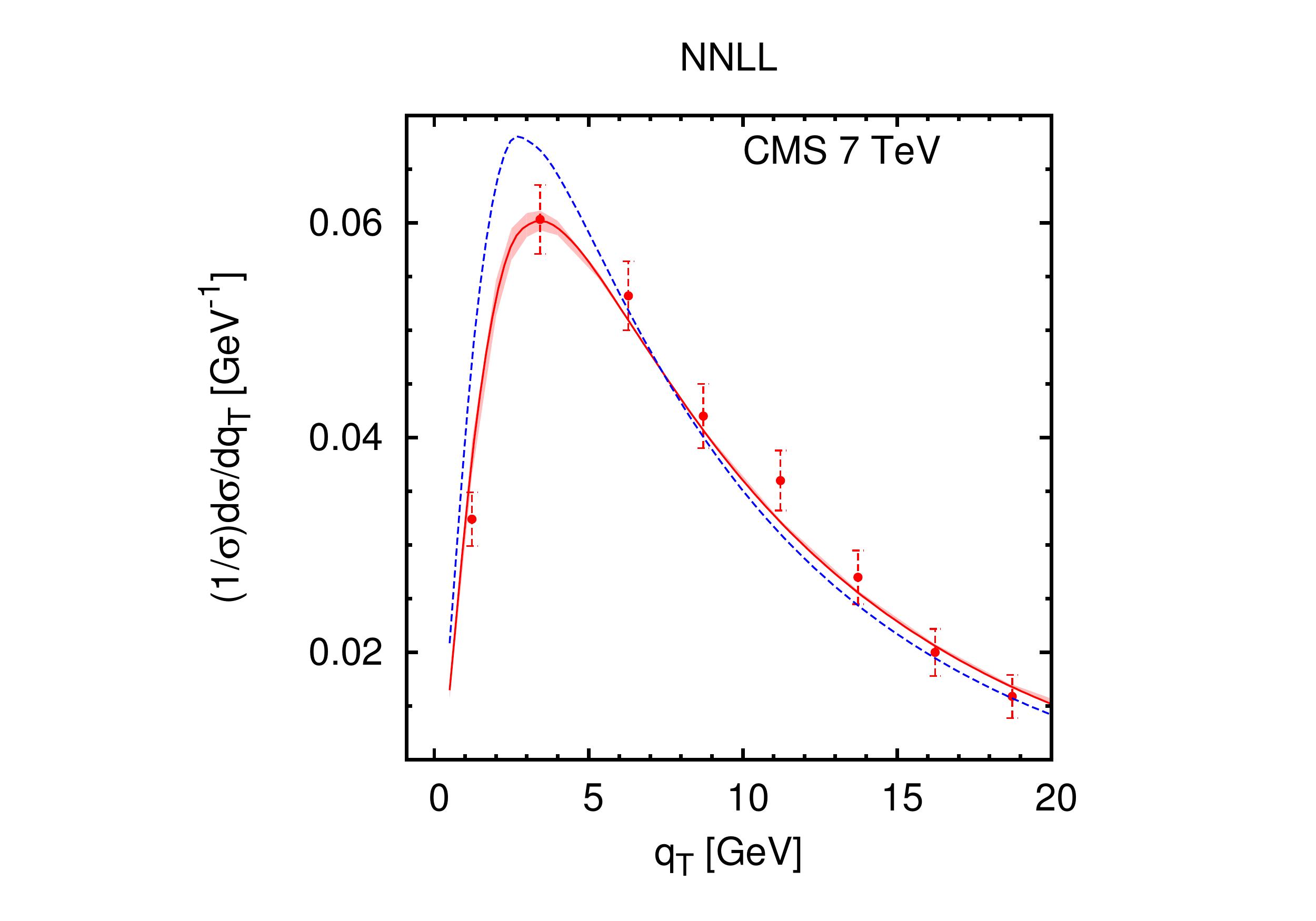}
%
\caption{The red solid curve is our prediction for $(1/\sigma)d\sigma/dq_T$ based on the global fit with $D^{\rm NP}=0$ (Eq.~(\ref{eq:FqN3})), \mbox{$Q_i=Q_0+q_T$}, at NNLL-NNLO accuracy compared to CMS experimental data~\cite{Chatrchyan:2011wt}. The band comes from the statistical error on the fitted parameter $\l_1$.
The blue dashed line is the full resummed result at NNLL-NNLO accuracy with no non-perturbative input, $\la_1=\la_2=0$.}
\label{fig-3}       
\end{figure}
\section{Conclusions}
\label{concl}

The TMD formalism is a powerful tool to analyze perturbative and non-perturbative effects in $q_T$ spectra.
In this talk we report the results for the fit of   the DY and $Z$-boson production data to fix the non-perturbative part of TMDPDFs.
In order to have a reliable fixing of the non-perturbative inputs one has to provide a fully resummed expression for the perturbative part. The fully resummed cross section in fact is less sensitive to the factorization scale dependence and this allows a more stable extraction of the non-perturbative pieces of the TMDs. To this aim,
we have performed a detailed and complete study of the perturbative inputs. In particular we have used the TMD evolution kernel at NNLL~\cite{Echevarria:2012pw} which, to our knowledge, was never used before in a global fit of this kind.
We have also discussed the matching of TMDPDFs onto PDFs with the exponentiation and fully resummation of the  corresponding coefficient. We argue that the exponentiated part of this matching coefficient is spin independent and should be included in the analysis of other types of TMDPDFs. This part is fundamental to have a reliable description of the TMDs both at NLL and NNLL accuracy.

One of the important aspects of our perturbative analysis is that the factorization scale is fixed in momentum space  instead of the more usual impact parameter space.
This choice provides a good stability of the perturbative series and offers a new understanding of the data and of the model dependence of the TMDs.
We find that the NNLL fit clarifies several issues about the non-perturbative nature of TMDs.

The model-dependent non-perturbative inputs for the TMDPDF are studied in order to minimize the number of non-perturbative parameters and to provide a good description of the data.
We find that the $Z$-boson data are better described by an exponential damping factor in impact parameter space rather than a Gaussian one. The associated parameter, called $\l_1$ in the text, has a stable value within the errors, which are mainly of statistical origin.
The low-energy data explore values of the impact parameter higher than those covered in the case of $Z$-boson production.
We find that a polynomial correction with a new parameter, called $\l_2$ in the text, plays a relevant role in this respect and both corrections, induced by $\l_{1,2}$, do not depend on the dilepton invariant mass $Q$.
The values of these parameters can be fixed by fitting data for DY and $Z$-boson production and the NNLL resummation greatly reduces the theoretical error on this determination.

Particular attention has been paid to the study of the $Q$ dependence of the non-perturbative model. The insertion of this contribution (parametrized by $\l_3$) provides only an improvement of the $\chi^2$ at the price of adding a new parameter to the fit.
Nevertheless, given the actual uncertainties on data and collinear PDFs, the need for this correction in the fits cannot be firmly established.
Increasing the precision of the experimental data can be crucial to fix this issue.
This aspect was completely unclear  in previous fits of the same data,  as in Ref.~\cite{Landry:2002ix}.
In that work the perturbative part of the evolution kernel is minimized with a particular choice of scales.
As a result the evolution kernel is completely  described by a model whose parameters are called $b_{\rm max}$ and $g_2$. The values  of these parameters extracted from the fit   are dependent on the energy scales of the
fit and  are not universal. Using different  data of Drell-Yan or DIS processes at different energy scales one expects different results for the  values of these parameters. This problem is expected to be largely reduced in our approach, because of the resummations of the evolution kernel.
Moreover a fully resummed evolution kernel, together with other exponentiations and resummations in the various matching coefficients that appear in the cross section, avoid an excessive use of a modelization of the cross section, making the predictions more stable.

We consider this work as a first step towards the proper understanding of non-perturbative effects in transverse momentum distributions. Several important perturbative pieces, recently calculated~\cite{Catani:2011kr,Catani:2012qa,Gehrmann:2012ze,Gehrmann:2014yya}, can be used in an approximate N$^3$LL analysis and will be included in a forthcoming study.

We point out that fixing the non-perturbative part of transverse momentum distributions can improve substantially the theoretical precision needed for the current LHC experiments, as our prediction for $Z$-boson $q_T$ spectrum at CMS shows in Fig.~\ref{fig-3}.  This picture shows that the differential cross section at the peak without non-perturbative inputs (dashed line) is not able to describe the data. The non-perturbative pieces fitted using low-energy data as well as Tevatron vector boson production data provide a prediction with very small uncertainty which agrees with CMS data, (see the band in Fig.~\ref{fig-3}).

Finally, we comment on the use of this formalism for SIDIS processes.
The parameters $\l_{1,2}$ are specific of the unpolarized TMDPDF and can be directly used also for this type of analysis, while different values of these parameters are expected for the fragmentation functions. On the other hand the parameter $\l_3$ is a universal correction and, as such, it is the same in DY and SIDIS processes.
In order to confirm the universality of the TMDPDFs (and their non-perturbative behaviour) in a future work we plan
to analyze SIDIS data adopting the parameters so extracted in the present study.

\section*{Acknowledgements}
U.D.~is grateful to the Department of Theoretical Physics II of the Universidad Complutense of Madrid for the kind hospitality extended to him since the earlier stages of this work.
M.G.E.~is supported by the ``Stichting voor Fundamenteel Onderzoek der Materie'' (FOM), which is financially supported by the ``Nederlandse Organisatie voor Wetenschappelijk Onderzoek'' (NWO).
U.D.~and S.M.~acknowledges support from the European Community under the FP7 program ``Capacities - Research Infrastructures'' (HadronPhysics3, Grant Agreement 283286). S.M.~is partly supported by the ``Progetto di Ricerca Ateneo/CSP'' (codice TO-Call3-2012-0103).
I.S.~is supported by the Spanish MECD grant, FPA2011-27853-CO2-02.

\end{document}